\documentclass[aps,prc,twocolumn,showpacs,superscriptaddress,footinbib]{revtex4-1}

\usepackage{graphicx}
\usepackage{dcolumn}
\usepackage{bm}
\usepackage{enumerate}
\usepackage{amsmath}

\usepackage[normalem]{ulem}  

\begin{document}


\title{SU(3) analysis for B(E2) anomaly}

\author{Yu-xin Cheng}
\affiliation{School of Physics, Liaoning University, Shenyang 110036, People's Republic of China}

\author{De-hao Zhao}
\affiliation{School of Physics, Liaoning University, Shenyang 110036, People's Republic of China}

\author{Yue-yang Shao}
\affiliation{College of Physics, Tonghua Normal University, Tonghua 134000, People's Republic of China}

\author{Li Gong}
\affiliation{School of Physics, Liaoning University, Shenyang 110036, People's Republic of China}

\author{Tao Wang}
\email{suiyueqiaoqiao@163.com}
\affiliation{College of Physics, Tonghua Normal University, Tonghua 134000, People's Republic of China}

\author{Xiao-shen Kang}
\email{kangxiaoshen@lnu.edu.cn}
\affiliation{School of Physics, Liaoning University, Shenyang 110036, People's Republic of China}

\date{\today}

\begin{abstract}
B(E2) anomaly is becoming a hot topic in the field of nuclear structure. Since the B(E2) anomaly was experimentally found, understanding the mechanism of its production has become an important problem. Theoretical studies have found that the SU3-IBM and other extended IBM theories can give explanations, but different mechanisms were found. In this paper, the concept ``SU(3) analysis'' for B(E2) anomaly is proposed, and some new conclusions are obtained. The SU(3) third-order interaction $[L\times Q \times L]^{(0)}$ and level-crossing are both vital for the emergence of the B(E2) anomaly.  This technique can help us to better understand the realistic reason of the B(E2) anomaly.
\end{abstract}

\maketitle

\section{Introduction}

50 years ago, the interacting boson model (IBM) was proposed by Arima and Iachello \cite{Iachello75,Iachello87}, which is an algebraic model for describing the collective excitations in nuclear structure. For the simplest case, only the $s$ and $d$ bosons with angular momentum $L=0$ and $L=2$ are considered to construct the Hamiltonian, which has the SU(6) symmetry. There exists four dynamical symmetry limits in this model: the U(5) symmetry limit (spherical shape), the SU(3) symmetry limit (prolate shape), the O(6) symmetry limit ($\gamma$-soft rotation) and  the $\overline{\textrm{SU(3)}}$ symmetry limit (oblate shape) \cite{Jolie01}. Thus shape phase transitions between different shapes can be also studied by this model \cite{Warner02,Casten06,Casten07,Bonatsos09,Casten09,Jolie09,Casten10,Jolos21,Fortunato21,Cejnar21,Jolie00,Cejnar03,Iachello04,Wang08}.

Although previous IBM is simple and consistent, some of the experimental anomalies seem to be indescribable by this model, such as the B(E2) anomaly \cite{Grahn16,Saygi17,Cederwall18,Goasduff19} and the Cd puzzle \cite{Garrett08,Garrett08,Garrett12,Batchelder12,Garrett19,Garrett20,Garrett18}. In the B(E2) anomaly, the ratio $E_{4/2}$ of the energies of the $4_{1}^{+}$, $2_{1}^{+}$ states is larger than 2.0 (a feature for the collective excitations), but the ratio $B_{4/2}$ of the E2 transitions $B(E2;4_{1}^{+}\rightarrow 2_{1}^{+})$ and $B(E2;2_{1}^{+}\rightarrow 0_{1}^{+})$ can be much smaller than 1.0 (a traditional signal for the non-collective behaviors). This seemingly conflicting anomaly cannot possibly be explained by previous theories in nuclear structure \cite{Grahn16,Saygi17,Cederwall18,Goasduff19}. In the Cd puzzle, the experimental data did not confirm the existence of the phonon excitations of the spherical nucleus \cite{Garrett08,Garrett08,Garrett12,Batchelder12,Garrett19,Garrett20}, instead questioned it \cite{Garrett18}. Experimentally, the B(E2) anomaly and the Cd puzzle may occur in adjacent nuclei, such as $^{72-76}$Zn \cite{Louchart13,Illana23,Hellgartner23} and even in a single nucleus, such as $^{114}$Te \cite{Moller05,Ray20}, so they may have a common origin.

Moreover, Otsuka \emph{et al.} found that nuclei previously considered as prolate shape should be rigid triaxial \cite{Otsuka19,Otsuka21,Otsuka}, making previous IBM descriptions not particularly convenient. In previous IBM, the spectra of the prolate shape (the SU(3) symmetry limit) and the oblate shape (the $\overline{\textrm{SU(3)}}$ symmetry limit) are the same \cite{Jolie01}, but this mirror symmetry can not be found in realistic nuclei \cite{Jolie03}.

Recently an extension of the interacting boson model with SU(3) higher-order interactions (SU3-IBM) was proposed by one of the authors (T.Wang) to resolve these various anomalies. In this model, the SU(3) symmetry plays the most important role and dominates all the quadrupole deformations. Through detailed discussions, it is found that this model can indeed better describe the collective behaviors of atomic nuclei, which is beyond researchers's expectations. It can describe the B(E2) anomaly and the Cd puzzle \cite{Wang22,Wang25}. Although the B(E2) anomaly can not be described by previous nuclear theories \cite{Grahn16,Saygi17,Cederwall18,Goasduff19}, there are many possibilities in the SU3-IBM and other extended IBM theories \cite{Wang20,Zhang22,Wangtao,Zhang24,Pan24,Zhang25,Zhang252}. In this paper, we further distinguish different mechanisms by the introduction of SU(3) analysis.  The spherical-like spectra for resolving the Cd puzzle were really found in $^{106}$Pd \cite{WangPd}. The SU3-IBM can be also used to explain the prolate-oblate shape asymmetric transitions in the Hf-Hg region \cite{Fortunato11,Zhang12,Wang23}, to describe the $\gamma$-soft behaviors in $^{196}$Pt at a better level \cite{WangPt,ZhouPt}, to describe the E(5)-like spectra in $^{82}$Kr \cite{Zhou23}, and to explain the unique boson number odd-even phenomenon in $^{196-204}$Hg \cite{WangHg} which was first found in  \cite{Zhang12}. Furthermore it can well describe the rigid triaxiality in $^{166}$Er \cite{ZhouEr}. Recently the shape phase transition from the new $\gamma$-soft phase to the prolate shape is also found and $^{108}$Pd is the critical nucleus \cite{Zhao251,Zhao252}. These results have overturned our traditional understanding of nuclear structure, because it has captured almost all the shape patterns and given new collective patterns for describing various anomalous behaviors.

In this paper, based on previous results \cite{Wang20,Zhang22,Wangtao,Zhang24,Pan24,Zhang25,Zhang252}, especially the four papers \cite{Wang20,Zhang22,Pan24,Zhang252}, we continue to discuss the mechanisms in the B(E2) anomaly. So far, there has existed many possible explanations for the B(E2) anomaly, which will require some more fundamental discussions to make some key parts clearer. We introduce the concept ``SU(3) analysis'', which is used to reanalyse the results obtained in these studies within the SU(3) symmetry limit. When the parameter in front of the SU(3) third-order interaction $[L\times Q \times L]^{(0)}$ evolves, the partial low-lying levels and the values of $B(E2;2_{1}^{+}\rightarrow 0_{1}^{+})$, $B(E2;4_{1}^{+}\rightarrow 2_{1}^{+})$ and $B(E2;6_{1}^{+}\rightarrow 4_{1}^{+})$ are discussed.  We find three new results: (1) The interaction $[L\times Q \times L]^{(0)}$ is critical for the SU(3) anomaly; (2) Level-crossing within the SU(3) symmetry limit is crucial for the SU(3) anomaly; (3) Not only the value of $B(E2;4_{1}^{+}\rightarrow 2_{1}^{+})$ but also the ones of $B(E2;6_{1}^{+}\rightarrow 4_{1}^{+})$ and $B(E2;2_{1}^{+}\rightarrow 0_{1}^{+})$ can be anomalous. Thus the SU(3) analysis is a useful tool for identifying the real cause. We expect it can be used in future discussions on the B(E2) anomaly. If not analyzed, the conclusions obtained may be incomplete or incorrect.

\section{SU3-IBM Hamiltonian}

In the SU3-IBM, only the U(5) symmetry limit and the SU(3) symmetry limit are included. In the SU(3) symmetry limit, the SU(3) second-order Casimir operator $-\hat{C}_{2}[SU(3)]$ can describe the prolate shape and the SU(3) third-order Casimir operator $\hat{C}_{3}[SU(3)]$ can describe the oblate shape, which fundamentally distincts from previous IBM \cite{Wang23}. The rigid triaxial shape can be obtained by the combinations of the square of the SU(3) second-order Casimir operator $\hat{C}_{2}^{2}[SU(3)]$ and the $-\hat{C}_{2}[SU(3)]$, $\hat{C}_{3}[SU(3)]$. Moreover, the SU(3) invariants $[L\times Q \times L]^{(0)}$ and $[(L \times Q)^{(1)}\times (L \times Q)^{(1)}]^{(0)}$ are necessary.

The Hamiltonian is as follows
\begin{eqnarray}
\hat{H}&=&\alpha\hat{n}_{d}+\beta\hat{C}_{2}[SU(3)]+\gamma\hat{C}_{3}[SU(3)]+\delta\hat{C}_{2}^{2}[SU(3)]   \nonumber\\
&&+\eta[\hat{L}\times \hat{Q} \times \hat{L}]^{(0)}+\zeta[(\hat{L}\times \hat{Q})^{(1)} \times (\hat{L} \times \hat{Q})^{(1)}]^{(0)} \nonumber\\
&&+\xi \hat{L}^{2}
\end{eqnarray}
where $\alpha$, $\beta$, $\gamma$, $\delta$, $\eta$, $\zeta$ and $\xi$ are seven fitting parameters. $\hat{Q}=[d^{\dag}\times\tilde{s}+s^{\dag}\times \tilde{d}]^{(2)}-\frac{\sqrt{7}}{2}[d^{\dag}\times \tilde{d}]^{(2)} $ is the SU(3) quadrupole operator. The first four interactions determine the quadrupole shapes of the ground state of the nucleus, and the positions of the $0^{+}$ states of the excited levels. The latter three ones are the dynamical interactions and can be used to change the features of the non-$0^{+}$ states.

For understanding the B(E2) anomaly, the B(E2) values are necessary. The $E2$ operator is defined as
\begin{equation}
\hat{T}(E2)=q\hat{Q},
\end{equation}
where $q$ is the boson effective charge. The evolutions of $B(E2; 2_{1}^{+}\rightarrow 0_{1}^{+})$, $B(E2; 4_{1}^{+}\rightarrow 2_{1}^{+})$, $B(E2; 6_{1}^{+}\rightarrow 4_{1}^{+})$ values are discussed.

\section{SU(3) analysis}

Although the B(E2) anomaly can not be explained by previous nuclear structure theories, in the extended IBM with higher-order interactions, it can be described by many ways \cite{Wang20,Zhang22,Wangtao,Zhang24,Pan24,Zhang25,Zhang252}. A key problem is that whether these explanations are related to the SU(3) symmetry. Obviously, this needs to consider these explanations in the SU(3) symmetry limit. For explaining the B(E2) anomaly, the boson number operator $\hat{n}_{d}$ is needed in some descriptions or the SU(3) quadrupole operator $\hat{Q}$ is replaced by the generalized quadrupole operator $\hat{Q}_{\chi}=[d^{\dag}\times\tilde{s}+s^{\dag}\times \tilde{d}]^{(2)}+\chi[d^{\dag}\times \tilde{d}]^{(2)}$. Thus for SU(3) analysis, the $\hat{n}_{d}$ interaction should be removed (here $\alpha$=0), and the  $\hat{Q}_{\chi}$ is replaced by the $\hat{Q}$ again (here $\chi=-\frac{\sqrt{7}}{2}$). If these operations are feasible, the SU(3) analysis exists, which is important for understanding the emergence of the B(E2) anomaly. This analysis has been ignored in previous studies, and in this paper, as we will see, it is very important.

B(E2) anomaly implies that $B_{4/2}<1.0$ while $E_{4/2}\geq 2.0$. These two quantities involve only $0_{1}^{+}$, $2_{1}^{+}$ and $4_{1}^{+}$ states, so they are important for the success of any nuclear structure theories. This point is similar to the magic number in atomic nuclei, which is only related to the $0_{1}^{+}$ and $2_{1}^{+}$ states. It should be noticed that this B(E2) anomaly was first observed theoretically by Y. Zhang \emph{et al.} in \cite{zhang14}, but it is not consistent with the experimental data in \cite{Grahn16,Saygi17,Cederwall18,Goasduff19}, in which the $B_{4/2}$ values are much smaller than 1.0. To date, we have realized that the B(E2) anomaly is just an important phenomenon as the magic numbers. If a theory cannot explain the B(E2) anomaly, it is insufficient.

Through many careful calculations, we find that the SU(3) cubic interaction $[L\times Q \times L]^{(0)}$ plays an important role in explaining the B(E2) anomaly. (Whether this relationship is unique is still unclear) Thus when discussing the problem, let the coefficient $\eta$ of this interaction change, and we will see the characteristics of the B(E2) anomaly.

\begin{figure}[tbh]
\includegraphics[scale=0.33]{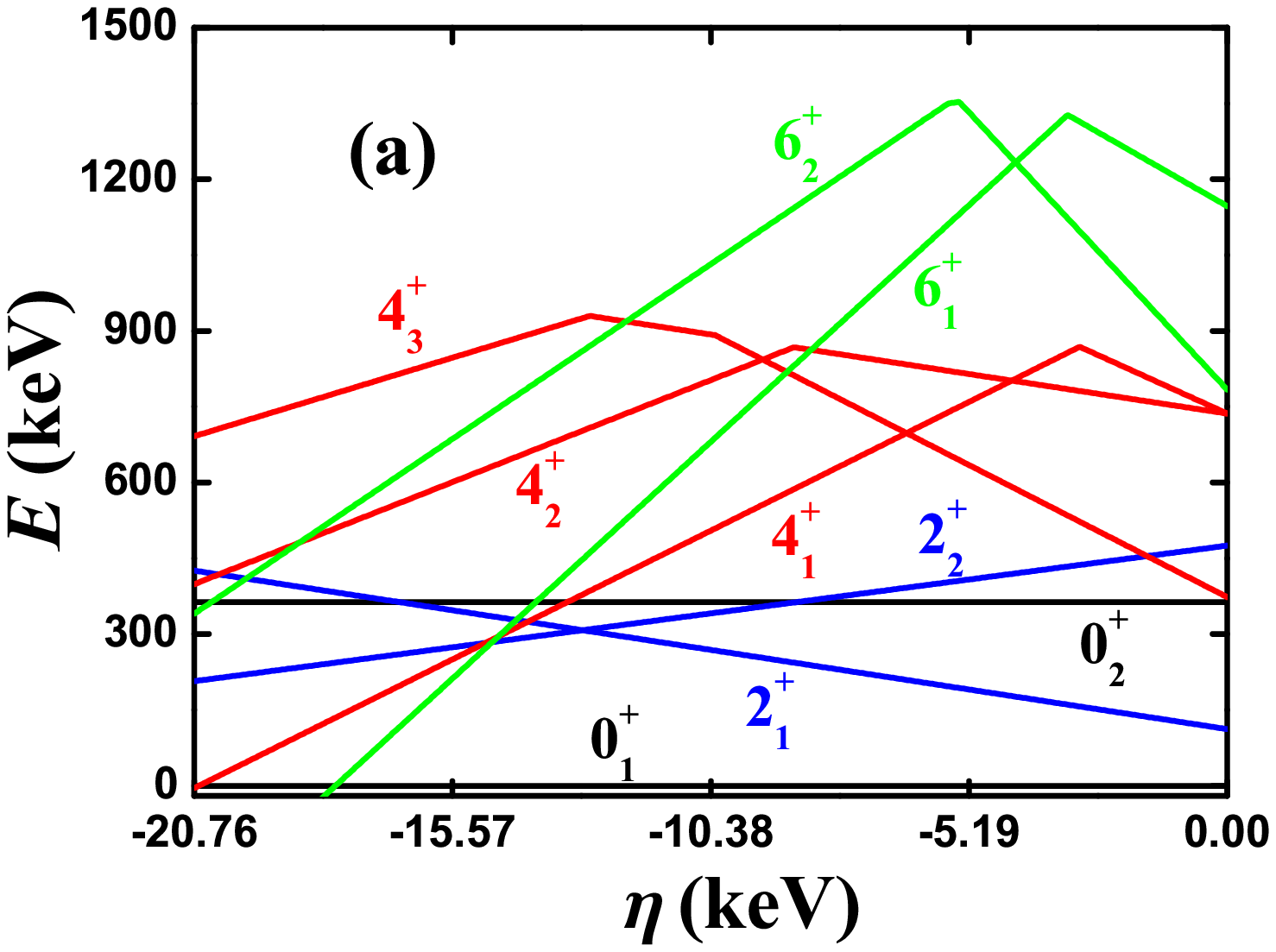}
\includegraphics[scale=0.33]{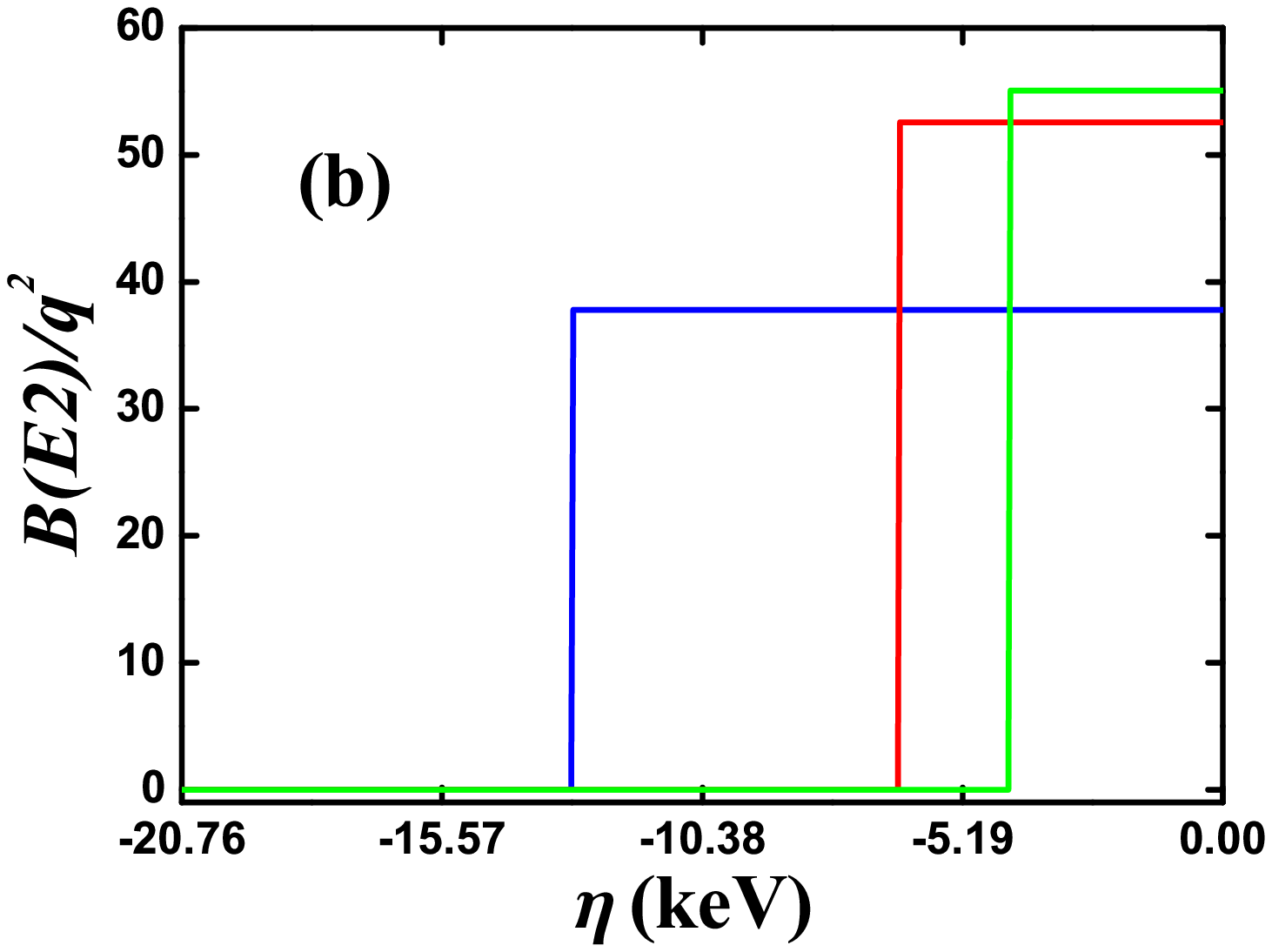}
\caption{(a) The evolutional behaviors of the partial low-lying levels as a function of $\eta$; (b) The evolutional behaviors of the $B(E2; 2_{1}^{+}\rightarrow 0_{1}^{+})$ (blue line) , $B(E2; 4_{1}^{+}\rightarrow 2_{1}^{+})$ (red line), $B(E2; 6_{1}^{+}\rightarrow 4_{1}^{+})$ (green line) as a function of $\eta$. The parameters are deduced from \cite{Wang20}.}
\end{figure}

\begin{figure}[tbh]
\includegraphics[scale=0.33]{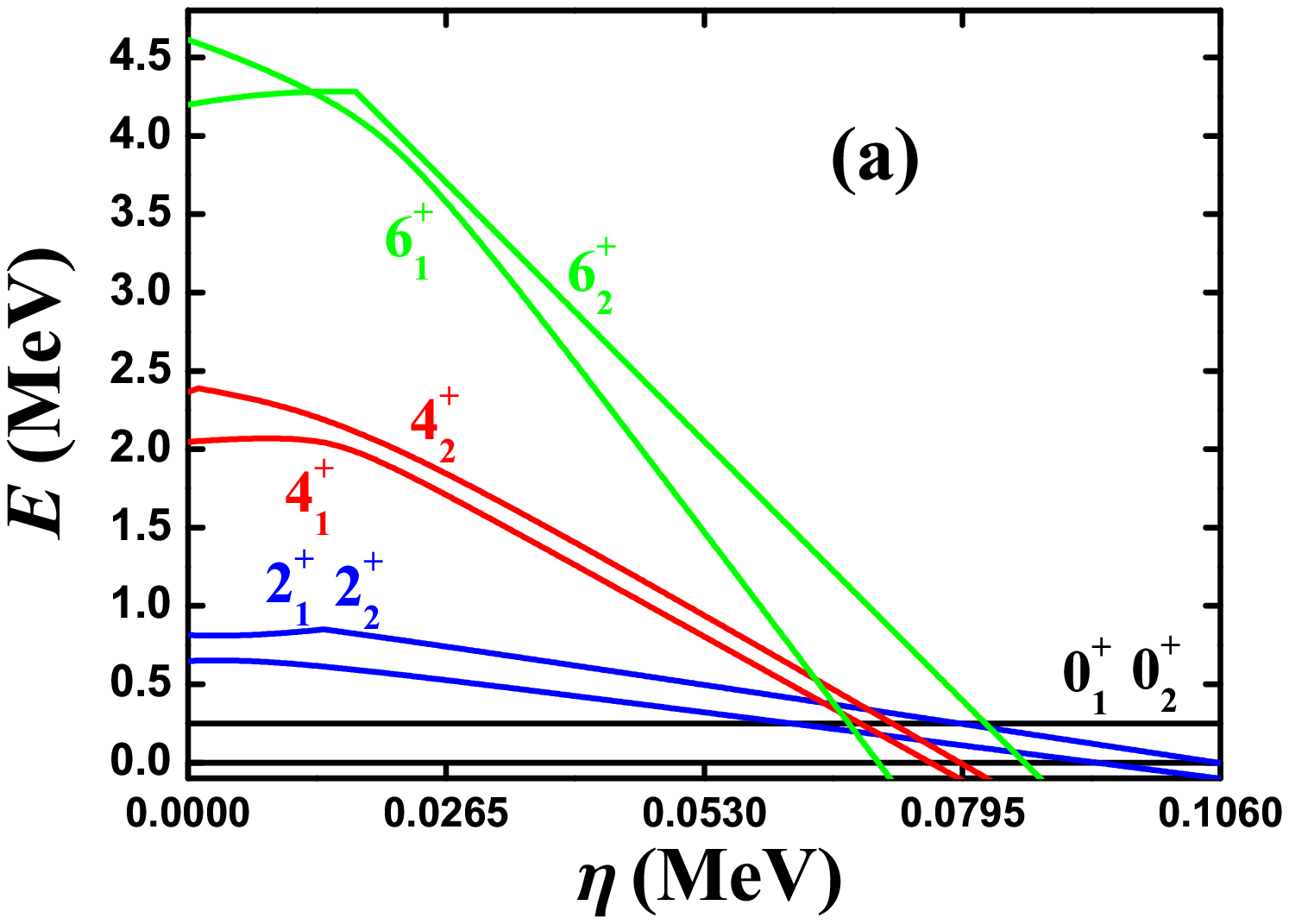}
\includegraphics[scale=0.33]{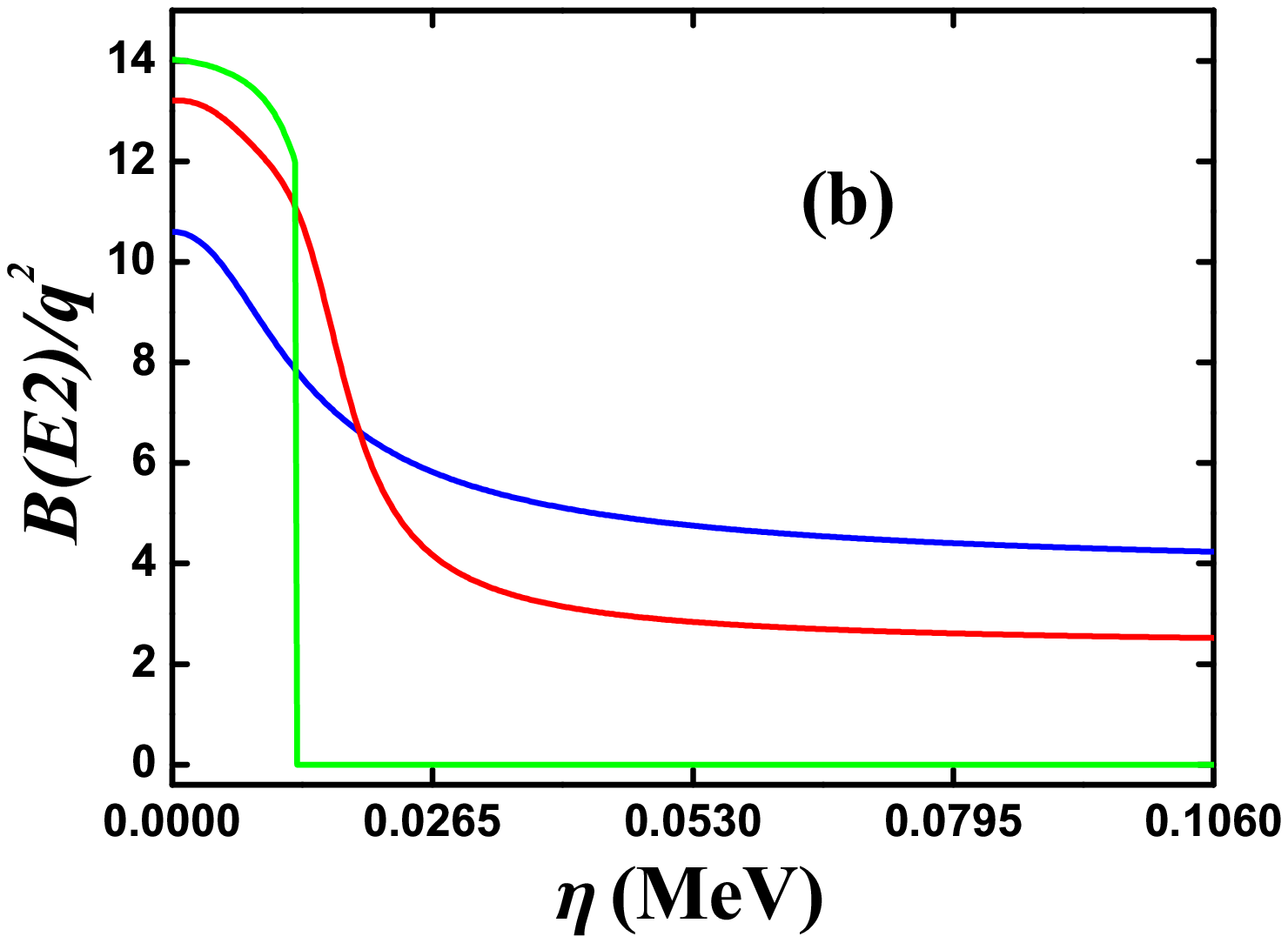}
\caption{(a) The evolutional behaviors of the partial low-lying levels as a function of $\eta$; (b) The evolutional behaviors of the $B(E2; 2_{1}^{+}\rightarrow 0_{1}^{+})$ (blue line), $B(E2; 4_{1}^{+}\rightarrow 2_{1}^{+})$ (red line), $B(E2; 6_{1}^{+}\rightarrow 4_{1}^{+})$ (green line) as a function of $\eta$. The parameters are deduced from \cite{Zhang22}.}
\end{figure}

\section{Analysis results}

Now we discuss the first example using SU(3) analysis. Fig. 1(a) shows the evolutional behaviors of the partial low-lying levels as a function of $\eta$ in the SU(3) symmetry limit in Ref. \cite{Wang20}. This is the first successful explanation for the B(E2) anomaly of realistic nucleus $^{170}$Os. The boson number is $N=9$. In \cite{Wang20}, the parameters are $\alpha=302.4$ keV, $\beta=-30.09$ keV, $\gamma=3.79$ keV, $\delta=0.0$ keV, $\eta=-10.38$ keV, $\zeta=0.0$ keV and $\xi=18.66$ keV. For SU(3) analysis, let $\alpha=0$. The ground state is a prolate shape with SU(3) irrep (18,0). In that paper, the two SU(3) third-order interactions are considered, but the two four-order interactions are not. From Fig. 1(a), we can see, when $\eta$ varies from 0 to -20.76 keV (the middle point is the parameter used in \cite{Wang20}), the line of the $4_{1}^{+}$ state can intersect with the one of one other $4^{+}$ state at $\eta=-6.45$ keV. In \cite{Wang20}, it is known that the $4^{+}$ state in the SU(3) irrep (10,4) becomes lower than the $4^{+}$ state in the SU(3) irrep (18,0).

However in Fig. 1(a), we can also notice some new features that are not observed by previous studies. The $6_{1}^{+}$ state can intersect with one other $6^{+}$ state at $\eta=-4.25$ keV and importantly, the $2_{1}^{+}$ state can intersect with one other $2^{+}$ state at $\eta=-12.97$ keV. Thus some new results can be found. In the SU(3) symmetry limit, if two states belong to different SU(3) irreps., the E2 transitions between them must be 0. In Fig. 1(b), the evolutional behaviors of the $B(E2; 2_{1}^{+}\rightarrow 0_{1}^{+})$ , $B(E2; 4_{1}^{+}\rightarrow 2_{1}^{+})$, $B(E2; 6_{1}^{+}\rightarrow 4_{1}^{+})$ as a function of $\eta$ are shown. We can see $B(E2; 4_{1}^{+}\rightarrow 2_{1}^{+})$ anomaly (here anomaly means the E2 transition is 0) can appear when $\eta\leq -6.45$ keV, and we can also notice that $B(E2; 6_{1}^{+}\rightarrow 4_{1}^{+})$ anomaly can happen when $\eta\leq -4.25$ keV and $B(E2; 2_{1}^{+}\rightarrow 0_{1}^{+})$ anomaly can occur when $\eta\leq -12.97$ keV. These anomalies are not mentioned in previous studies, and importantly we find they really exist in realistic nuclei, such as $B(E2; 6_{1}^{+}\rightarrow 4_{1}^{+})$ anomaly for $^{72}$Zn \cite{Louchart13} and $B(E2; 2_{1}^{+}\rightarrow 0_{1}^{+})$ anomaly for $^{166}$Os \cite{Stolze21}. We believe that the $B(E2; 2_{1}^{+}\rightarrow 0_{1}^{+})$ anomaly in $^{166}$Os is critical for understanding the reason of the B(E2) anomaly in realistic nuclei.

Clearly SU(3) analysis is a powerful technique to understand the B(E2) anomaly. Due to within the SU(3) symmetry limit, the level-crossing phenomena can occur which is induced by the $[L\times Q \times L]^{(0)}$ interaction. If the two levels belong to different SU(3) irrep., the E2 transition must be 0. This case is different from the SU(3) corresponding of the rigid triaxial description found in \cite{zhang14}.

\begin{figure}[tbh]
\includegraphics[scale=0.33]{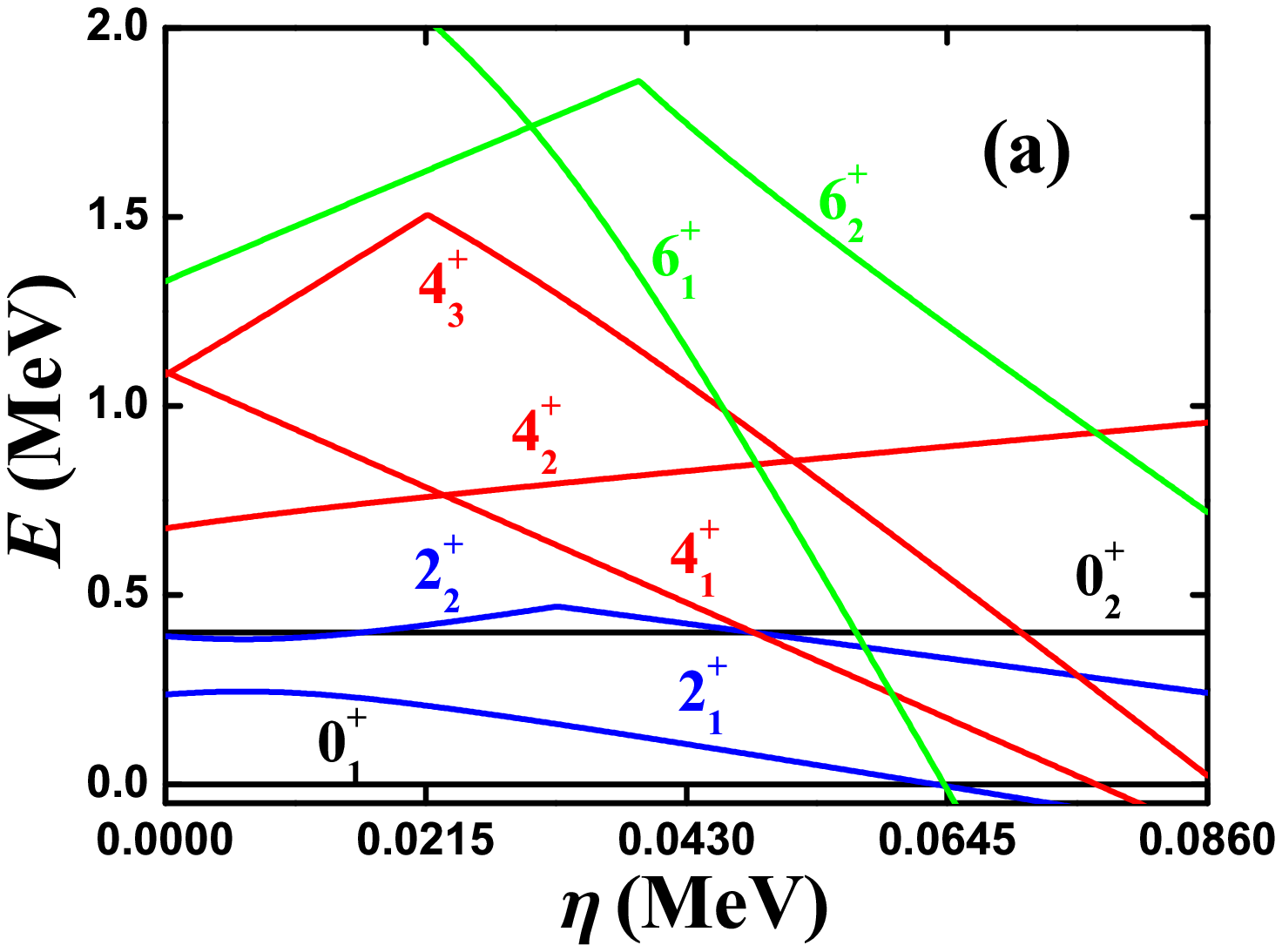}
\includegraphics[scale=0.33]{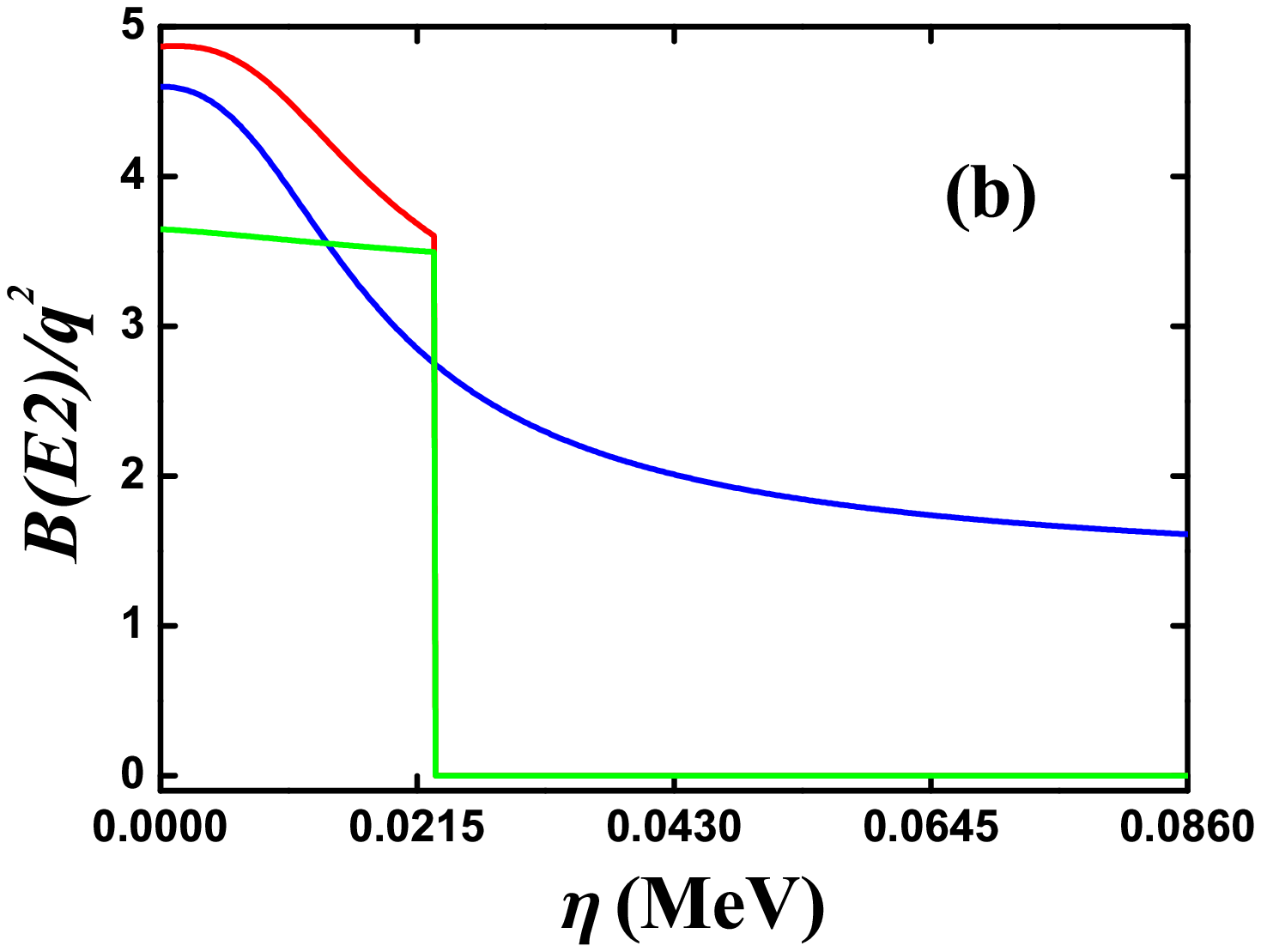}
\caption{(a) The evolutional behaviors of the partial low-lying levels as a function of $\eta$; (b) The evolutional behaviors of the $B(E2; 2_{1}^{+}\rightarrow 0_{1}^{+})$ (blue line), $B(E2; 4_{1}^{+}\rightarrow 2_{1}^{+})$ (red line), $B(E2; 6_{1}^{+}\rightarrow 4_{1}^{+})$ (green line) as a function of $\eta$. The parameters are deduced from \cite{Zhang22}.}
\end{figure}

Y. Zhang \emph{et al.} continued the rigid triaxial description in \cite{Zhang22}. Here we use $^{168}$Os for SU(3) analysis. The SU(3) corresponding of the rigid triaxial description was found in \cite{Draayer87,Draayer881,Draater882}, then it was used in the IBM to remove the degeneracy of the $\beta$ and $\gamma$ bands \cite{Isacker85} and to describe the rigid triaxial spectra \cite{Isacker00}. A key step is that Fortunato \emph{et al.} discussed the extended cubic $Q$-consistent Hamiltonian and found a new evolutional path from the prolate shape to the oblate shape \cite{Fortunato11}. This asymmetric shape evolution was also studied analytically by Y. Zhang \emph{et al.} \cite{Zhang12}. In \cite{zhang14}, Y. Zhang \emph{et al.} first found the B(E2) anomaly theoretically in studying the rigid triaxial rotor. Unfortunately, they don't realize that realistic nuclei can really show such features. The feature in \cite{zhang14} is that, when the angular moment $L$ in the ground band increases, the E2 transitional values $B(E2; L_{1}^{+}\rightarrow (L-2)_{1}^{+})$ really decrease, but at the beginning it decreases slowly.

\begin{figure}[tbh]
\includegraphics[scale=0.33]{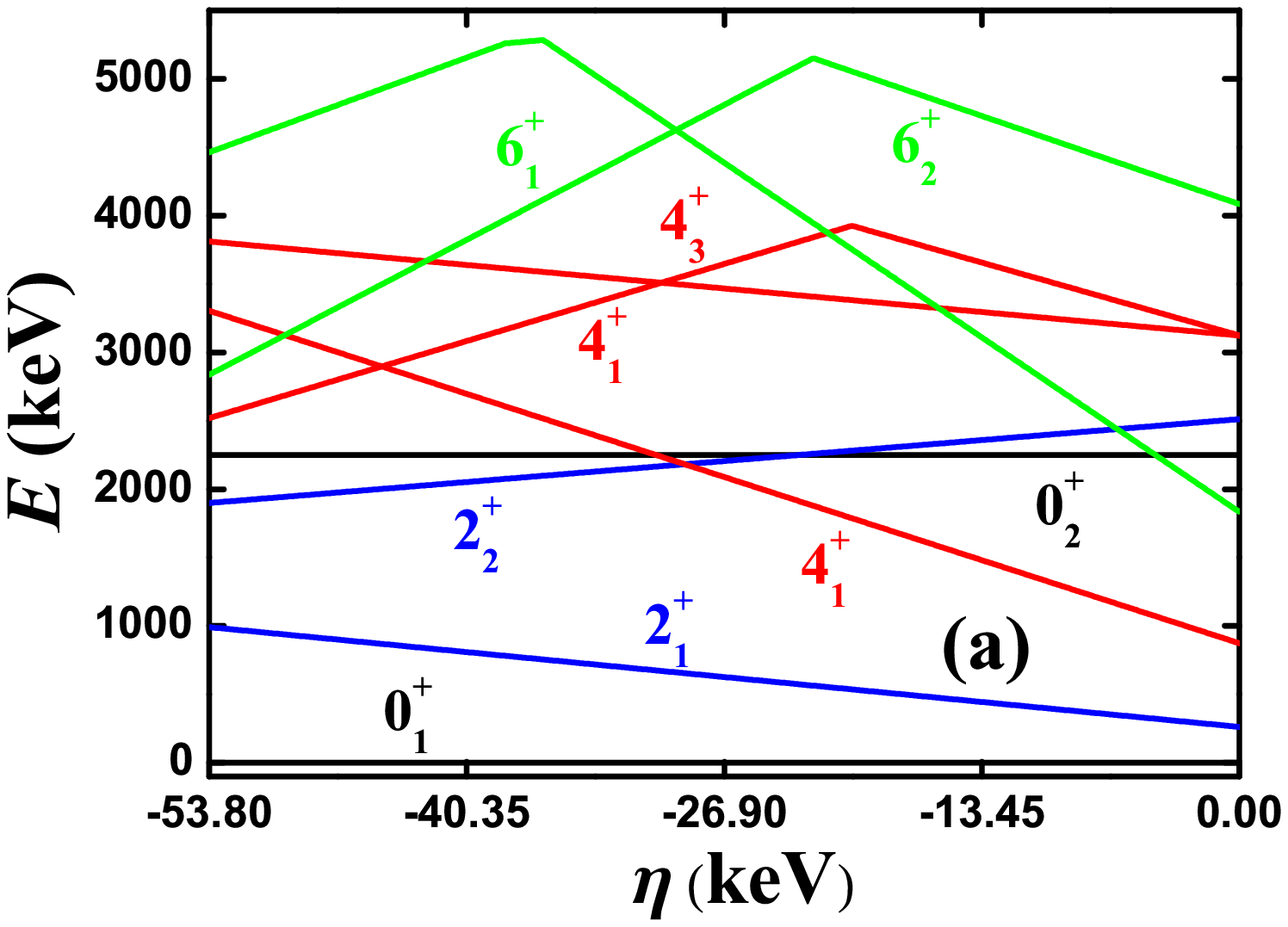}
\includegraphics[scale=0.33]{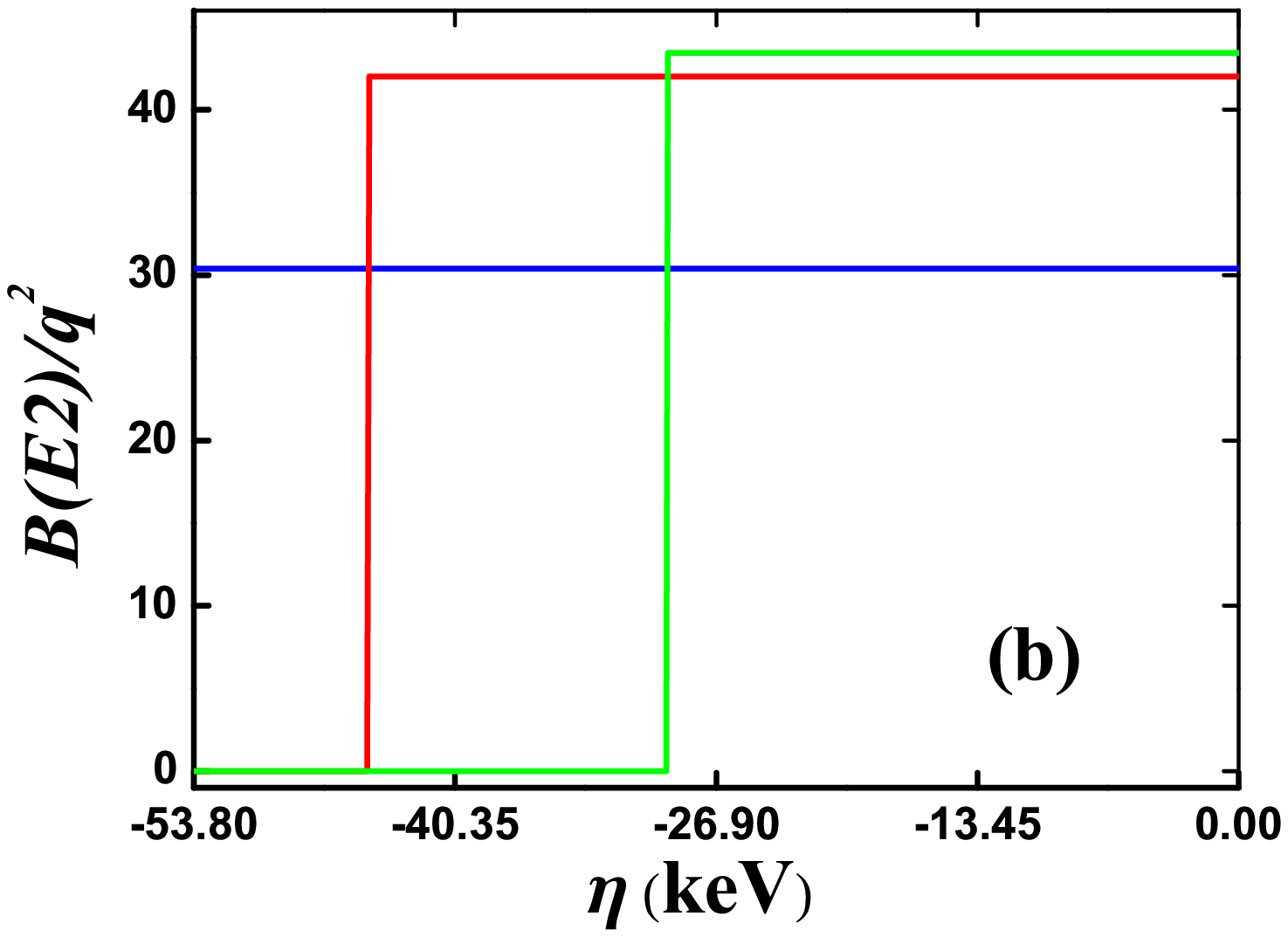}
\caption{(a) The evolutional behaviors of the partial low-lying levels as a function of $\eta$; (b) The evolutional behaviors of the $B(E2; 2_{1}^{+}\rightarrow 0_{1}^{+})$ (blue line), $B(E2; 4_{1}^{+}\rightarrow 2_{1}^{+})$ (red line), $B(E2; 6_{1}^{+}\rightarrow 4_{1}^{+})$ (green) as a function of $\eta$. The parameters are deduced from \cite{Pan24}.}
\end{figure}

For $^{168}$Os, the boson number is $N=8$. The parameters in Fig. 2 deduced from \cite{Zhang22} are $\alpha=0.022$ MeV, $\beta=0.096$ MeV, $\gamma=0.0027$ MeV, $\delta=0.0003$ MeV, $\eta=0.053$ MeV, $\zeta=-0.0045$ MeV and $\xi=0.094$ MeV. For SU(3) analysis, let $\alpha=0$. Fig. 2(a) shows the evolutional behaviors of the partial low-lying levels as a function of $\eta$ from 0 to 0.106 MeV, and the middle point is the parameter in \cite{Zhang22}. Obviously, the $4_{1}^{+}$ state really can not intersect with other $4^{+}$ states, but we can see that the $6_{1}^{+}$ state intersects with one other $6^{+}$ state. As shown in Fig. 2(b), the $B(E2; 4_{1}^{+}\rightarrow 2_{1}^{+})$ value can be lower than the $B(E2; 2_{1}^{+}\rightarrow 0_{1}^{+})$ value, so the B(E2) anomaly exists and results from the rigid triaxial rotor effect. However this ratio $B_{4/2}$ is 0.6, larger than the experimental value 0.34. Whether the rigid triaxial rotor effect can provide such a small $B_{4/2}$ value is a problem, which should be studied in future. Clearly, the $B(E2; 6_{1}^{+}\rightarrow 4_{1}^{+})$ anomaly results from the level-crossing effect. Here the SU(3) irrep of the ground state is (4,6), thus the B(E2) values are smaller than the ones in Fig. 1(b) if they exist.

\begin{figure}[tbh]
\includegraphics[scale=0.33]{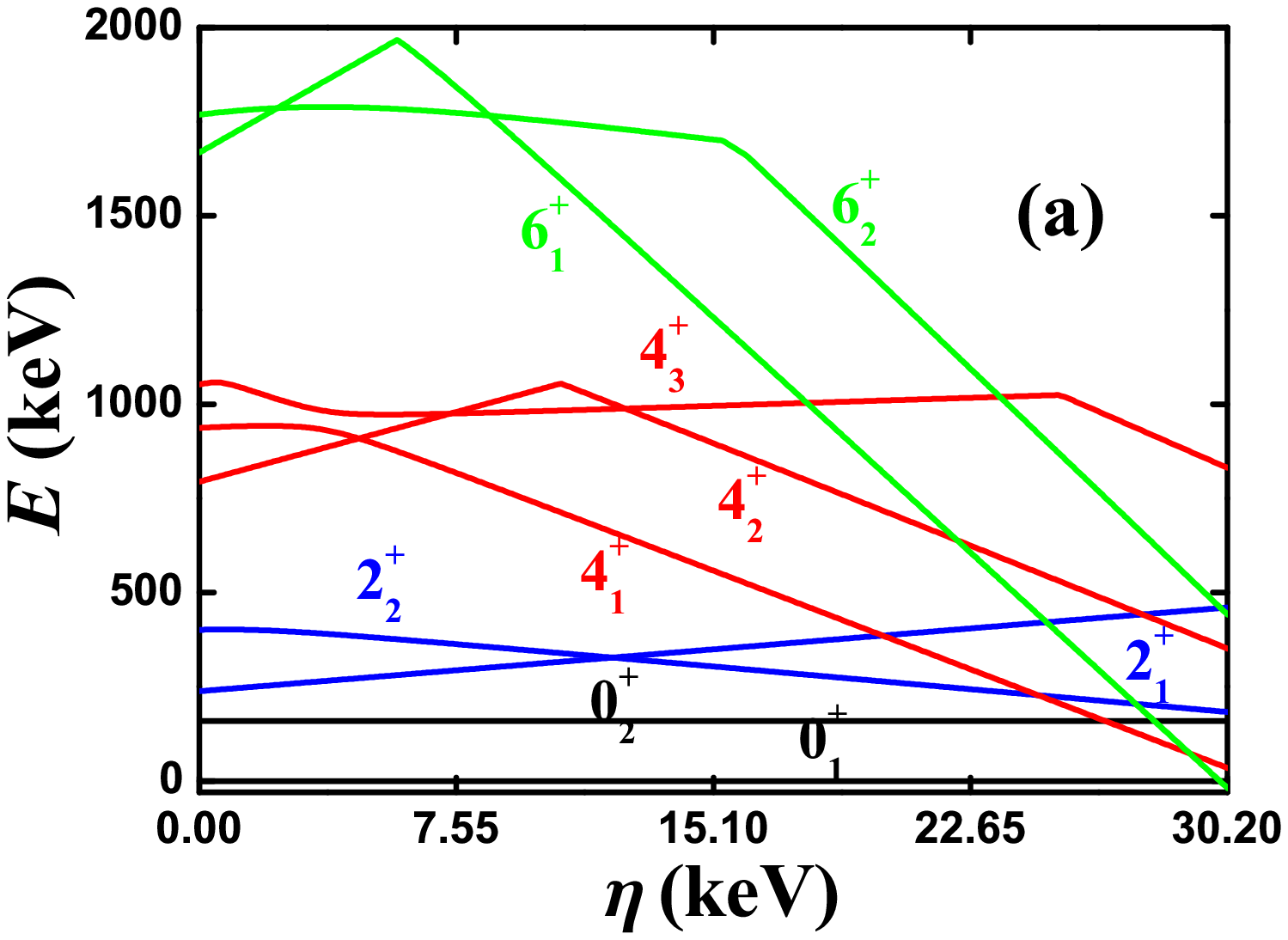}
\includegraphics[scale=0.33]{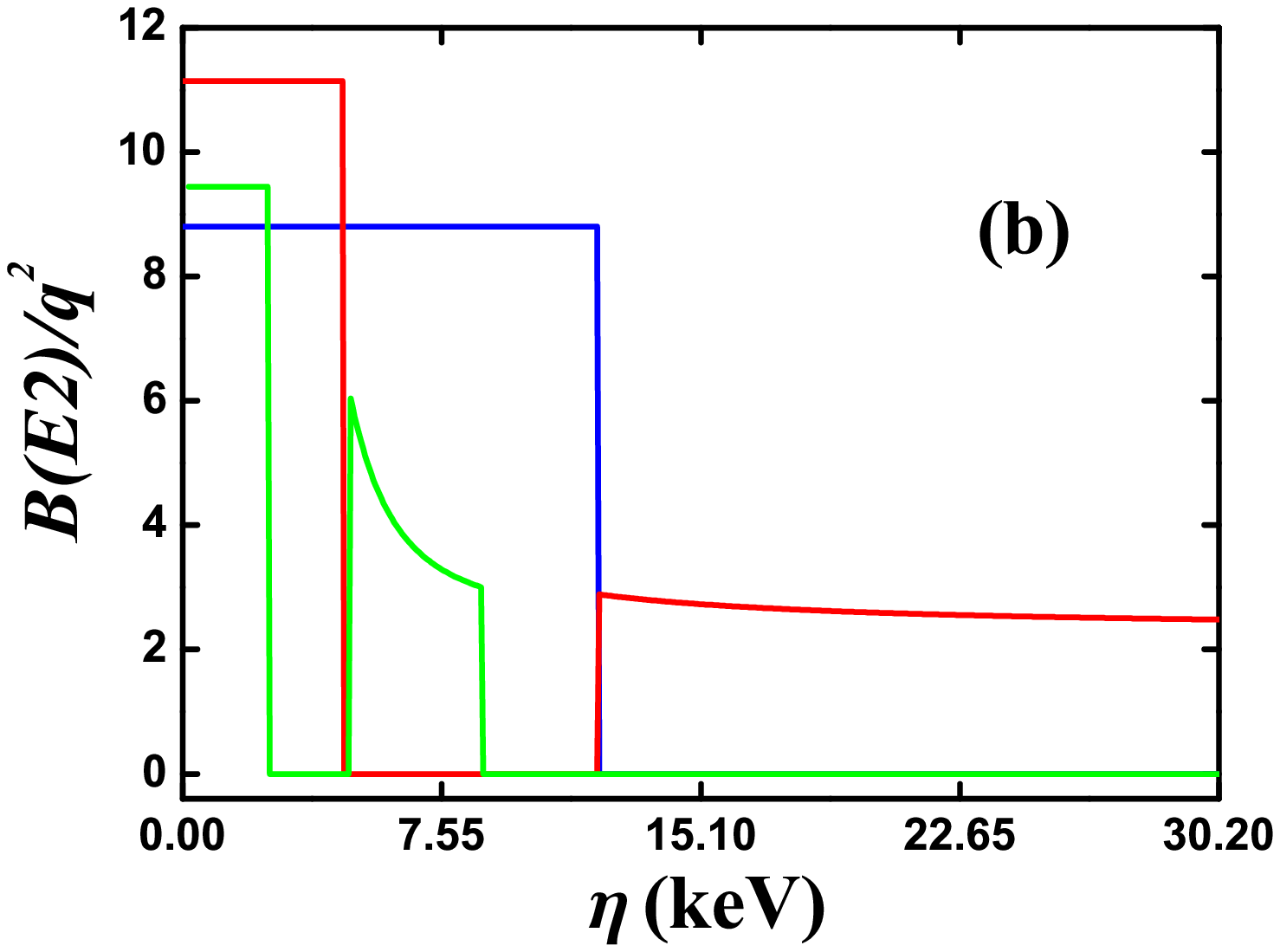}
\caption{(a) The evolutional behaviors of the partial low-lying levels as a function of $\eta$; (b) The evolutional behaviors of the $B(E2; 2_{1}^{+}\rightarrow 0_{1}^{+})$ (blue line), $B(E2; 4_{1}^{+}\rightarrow 2_{1}^{+})$ (red line), $B(E2; 6_{1}^{+}\rightarrow 4_{1}^{+})$ (green line) as a function of $\eta$. The parameters are deduced from \cite{Zhang252}.}
\end{figure}

In \cite{Zhang22}, there are another group of parameters for $^{168}$Os, which are $\alpha=0.099$ MeV, $\beta=0.0504$ MeV, $\gamma=0.0035$ MeV, $\delta=0.00055$ MeV, $\eta=0.043$ MeV, $\zeta=-0.0097$ MeV and $\xi=0.026$ MeV. For SU(3) analysis, let $\alpha=0$. Fig. 3(a) shows the evolutional behaviors of the partial low-lying levels as a function of $\eta$ from 0 to 0.086 MeV, and the middle point is the parameter in \cite{Zhang22}. Obviously, the $4_{1}^{+}$ state intersects with one other $4^{+}$ state at $\eta=0.023$ MeV and the $6_{1}^{+}$ state intersects with one other $6^{+}$ state at $\eta=0.0301$ MeV, which is larger than the one of the $4^{+}$ states crossing point. Thus this B(E2) anomaly results from level-crossing effect. However these crossing effects may be very complicated. Thus confirming the actual crossing effects of the experimental results requires more experimental data. If so, the emergence of the B(E2) anomaly may be an accidental effect. In Fig. 3(b) the $B(E2; 4_{1}^{+}\rightarrow 2_{1}^{+})$ anomaly and the $B(E2; 6_{1}^{+}\rightarrow 4_{1}^{+})$ anomaly can occur simultaneously. Here the SU(3) irrep of the ground state is (2,4), so the B(E2) values becomes smaller if they exist.

In previous IBM, O(5) symmetry exists when evolving from the U(5) limit to the O(6) limit, which results in the crossover of the $0_{2}^{+}$ and $0_{3}^{+}$ states. When the parameters deviate a little, significant energy level exclusion can occur. No doubt this is important to confirm such phenomena. Recently in the SU3-IBM, similar result was also found in \cite{Zhou23}, which is a level-anticrossing phenomenon.

Some important new results has been obtained recently. In \cite{Pan24}, the SU(3) quadrupole operator $\hat{Q}$ is replaced by the generalised quadrupole operator $\hat{Q}_{\chi}$. We also take the $^{168}$Os as an example, where $\alpha=0$, and other parameters are $\beta=-25.0$ keV, $\gamma=0$, $\delta=0$, $\eta=-26.9$ keV, $\zeta=0$ and $\xi=43.65$ keV and $\chi=-0.39$. For SU(3) analysis, let $\chi=-\frac{\sqrt{7}}{2}$. Here the SU(3) irrep of the ground state is (16,0) with the prolate shape. Fig. 4(a) shows the evolutional behaviors of the partial low-lying levels as a function of $\eta$ from 0 to -53.8 keV, and the middle point is the parameter in \cite{Pan24}. Intuitively, this evolutional behavior is very similar to the ones in Fig. 1(a). The $4_{1}^{+}$ state intersects with one other $4^{+}$ state at $\eta=-44.8$ keV, and the $6_{1}^{+}$ state also intersects with one other $6^{+}$ state at $\eta=-29.4$ keV. It implies that this B(E2) anomaly is also related to the SU(3) symmetry. We can notice that, the middle point shows a normal $B_{4/2}$ value, but it is near the anomalous region. This is very interesting, and gives a new mechanism for B(E2) anomaly, which is related with the level-anticrossing phenomenon. The B(E2) anomaly in the SU(3) symmetry limit and the B(E2) anomaly along the transitional region from the SU(3) symmetry limit to the O(6) symmetry limit are closely related. Detailed discussions can be seen in \cite{Li25}.

Similar phenomenon like in \cite{Pan24} can also appear in the SU3-IBM, which is found in \cite{Zhang252}. We also take $^{168}$Os for an example. Fig. 5(a) shows the evolutional behaviors of the partial low-lying levels as a function of $\eta$ from 0 to -53.8 keV, and the middle point is the parameter in \cite{Zhang252}. Here the SU(3) irrep of the ground state is (0,8) with the oblate shape. We can see that the $2_{1}^{+}$ and $2_{2}^{+}$ states intersects with each other at $12.1$ keV, the $4_{1}^{+}$ and $4_{2}^{+}$ states intersects with each other at $4.6$ keV, and the $6_{1}^{+}$ and $6_{2}^{+}$ states at $\eta=2.5$ keV. At the middle point, the $B_{4/2}$ value is infinity. This is very interesting. When the $d$ boson number operator is added, the B(E2) anomaly can be also found. This is also a level-anticrossing phenomenon, which will be discussed in another paper \cite{Tie25}. We can also confirm that this B(E2) anomaly is related to the SU(3) symmetry.

These results broaden our understanding of the B(E2) anomaly. Even the conditions, that (1) the $B(E2; 2_{1}^{+}\rightarrow 0_{1}^{+})$ value exists and (2) the $B(E2; 4_{1}^{+}\rightarrow 2_{1}^{+})$ is 0, are not satisfied in the SU(3) limit, the B(E2) can also appear.

If an explanation for B(E2) anomaly can have a SU(3) symmetry limit, the SU(3) analysis should be performed. From the above examples from \cite{Wang20,Zhang22,Wangtao,Zhang24,Pan24,Zhang25,Zhang252}, we can get many new results. There are some phenomena that seem to have nothing to do with the SU(3) symmetry, but are actually related to do with it. This may be a more complicated mechanism, which we need to further elaborate on. In the SU(3) analysis, any level-crossing can not occur but the B(E2) anomaly can happen, thus it can be considered from the rigid triaxial rotor effect. If any B(E2) anomaly does not exist in the SU(3) analysis but the B(E2) anomaly can happen when the $d$ boson number operator is added (this is almost impossible) or the $\hat{Q}$ is replaced by the $\hat{Q}_{\chi}$, this will be very important. A general discussion will be given in future with the extended $Q$-consistent Hamiltonian with up to fourth-order interactions. In a previous paper \cite{Wang22}, one of the authors (T. Wang) has proved that, in this extended Hamiltonian, when $\chi=0$ that is the $O(6)$ symmetry limit, the B(E2) anomaly can not happen. Thus it is important to discuss the evolving region from the SU(3) symmetry limit to the O(6) symmetry limit.

\section{Conclusion}

In this paper, we propose a powerful technique for understanding the B(E2) anomaly. It is the SU(3) analysis. From the discussions of the examples in \cite{Wang20,Zhang22,Wangtao,Zhang24,Pan24,Zhang25,Zhang252}, many new results are presented. There are three of the most important ones. First, the SU(3) third-order interaction   $[L\times Q \times L]^{(0)}$ is critical for the SU(3) anomaly. Whether this relationship is unique requires further investigation. Second, when this interaction is added, various level-crossing phenomena can happen. The causes of the B(E2) anomaly in realistic experiments require more experimental researches to determine. Looking for more mechanisms for the B(E2) anomaly also seems to be needed. Third, not only the value of $B(E2;4_{1}^{+}\rightarrow 2_{1}^{+})$ but also the $B(E2;6_{1}^{+}\rightarrow 4_{1}^{+})$ and $B(E2;2_{1}^{+}\rightarrow 0_{1}^{+})$ can be anomalous. It is very important to find more and more B(E2) anomaly results in the experiments.

In these SU(3) analysis, some new results can be obtained. These new mechanisms are related to the SU(3) symmetry, but at a more complicated level. We find they are relevant to the level-anticrossing phenomenon \cite{Li25,Tie25}. These works will help us to get a unified theoretical framework for explaining the B(E2) anomaly.

\end{document}